# LEDs for Energy Efficient Greenhouse Lighting

Devesh Singh, Chandrajit Basu, Merve Meinhardt-Wollweber, and Bernhard Roth

Hannover Centre for Optical Technologies, Nienburger Str. 17, 30167 Hannover, Germany

e-mail address: c.basu@hot.uni-hannover.de

**Abstract**

Light energy is an important factor for plant growth. In regions where the natural light source (solar radiation) is not sufficient for growth optimization, additional light sources are being used. Traditional light sources such as high pressure sodium lamps and other metal halide lamps are not very efficient and generate high radiant heat. Therefore, new sustainable solutions should be developed for energy efficient greenhouse lighting. Recent developments in the field of light source technologies have opened up new perspectives for sustainable and highly efficient light sources in the form of LEDs (light-emitting diodes) for greenhouse lighting. This review focuses on the potential of LEDs to replace traditional light sources in the greenhouse. In a comparative economic analysis of traditional vs. LED lighting, we show that the introduction of LEDs allows reduction of the production cost of vegetables in the long-run (several years), due to the LEDs' high energy efficiency, low maintenance cost and longevity. In order to evaluate LEDs as a true alternative to current lighting sources, species specific plant response to different wavelengths is discussed in a comparative study. However, more detailed scientific studies are necessary to understand the effect of different spectra (using LEDs) on plants physiology. Technical innovations are required to design and realize an energy efficient light source with a spectrum tailored for optimal plant growth in specific plant species.

# Introduction

Solid state lighting using light-emitting diode (LED) technology represents a fundamentally different and energy efficient approach for the greenhouse industry that has proficient advantages over gaseous discharge-type lamps (high pressure sodium lamps) currently used in most greenhouses [1, 2]. LED is a type of semiconductor diode which allows the control of spectral composition and the adaptation of light intensity to be matched to the plant photoreceptors in order to furnish better growth and to influence plant morphology as well as different physiological processes such as flowering and photosynthetic efficiency [3]. LEDs have the ability to produce high luminous flux with low radiant heat output and maintain their light output efficacy for years. The incandescent or fluorescent bulbs contain filaments that must be periodically replaced and consume a lot of electrical power while generating heat [4]. LEDs, however, do not have filaments and, thus, do not burn like incandescent or fluorescent bulbs. Due to low radiant heat production, LEDs can be placed close to plants and can be configured to emit high light fluxes even at high light intensities [4, 5].

An LED is a solid state device and can easily be integrated into digital control systems facilitating complex lighting programs such as varying intensity or spectral composition over a course of plant developmental stages [3]. Light under which plants are grown affects their growth and physiology (flowering and photosynthetic efficiency) in a complicated manner [6]. Light quality and quantity affect the signalling cascade of specific photoreceptors (phytochromes, cryptochromes and phototropins) which change the expression of a large number of genes. Using LEDs as a lighting source, it is possible not only to optimize the spectral quality for various plants and different physiological processes, but also to create a digitally controlled and energy efficient lighting system [7, 8].

The high capital cost of LED lighting systems is an important aspect delaying the establishment of LED technology in greenhouse lighting. However, technological development and mass production (based on high demand in general and in the greenhouse industry in future) is expected to reduce the capital and operating cost in the future significantly [2, 9, 10]. A properly designed LED light system can provide highly efficient performance and longevity well beyond any traditional lighting source [11]. Research on LED lighting for plant growth has been going on for almost two decades now. LED lighting on various vegetables has shown good results in terms of maximal productivity and optimal nutritional quality, paving the way for a wider acceptance of LED technology in greenhouse lighting in future. This review provides a summary of research done on plants (photosynthesis, growth, nutritional value and flowering) using LED lighting systems and addresses the important questions such as:

- Why should LED lighting systems be preferred over traditional lighting sources?
- What spectral composition should be used and should it be adjustable?
- What are the major challenges for LED lighting systems?

**LEDs and their practical perspectives**

Energy is an important factor which contributes about 20-30% of total production cost in greenhouse industry [12, 13]. Appropriate crop lighting is a necessity of the greenhouse industry, particularly in regions where the seasonal photoperiod (natural day length) fluctuates and there is not sufficient light for optimal plant growth. Nowadays, High Pressure Sodium (HPS) lamps are the most commonly used light sources in the greenhouse industry. HPS lamps operate at high temperature (≥200˚C), resulting in significant radiant heat emission (infrared) in the direct environment [14]. As a result HPS lamps cannot be placed close to plants and an ample ventilation system should be available to avoid too high temperatures close to the plants. This characteristic (radiant heat production) restricts the possibilities for future use of HPS lamps in energy efficient greenhouse concepts [15]. Thus, a new technology which significantly reduces the electricity consumption and produces low radiant heat for crop lighting while maintaining or improving the crop value (growth and nutritional value) is of great interest to the greenhouse industry.

LEDs represent an energy efficient approach for greenhouse lighting that has technical advantages over traditional light sources with fragile filaments, electrodes, or gas-filled pressurized lamp enclosures [11]. LEDs have great potential to play a variety of roles in greenhouse lighting. They are also well suited for research applications (e.g., in growth chambers for tissue culture applications). LEDs are solid state light emitting devices. The key structure of an LED consists of the chip (light-emitting semiconductor material), a lead frame where the die is placed and the encapsulation which protects the die (Fig. 1) [3]. Note that LEDs are available in different sizes and packages. An example of chip on board (COB) design is shown in Fig. 2. LEDs can be manufactured to emit broad-band (white) light or narrow-spectrum (colored) wavelengths specific to desired applications, for example plant responses [16]. In LEDs, waste heat is passed up separately from light-emitting surfaces through active heat sinks. This is particularly important for high intensity LEDs because the light source can be placed close to crop surfaces without risk of overheating and stressing the plants [11].

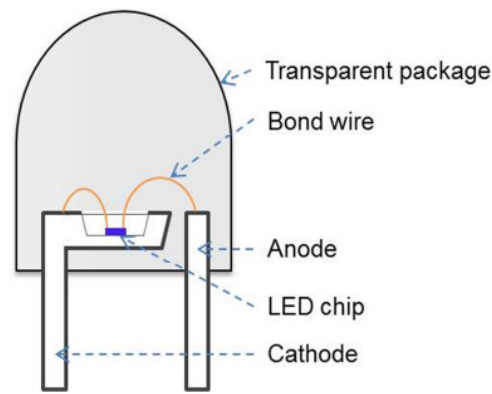

**Fig. 1** The key structure of an LED.

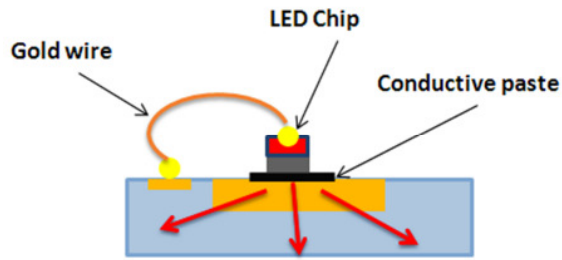

**Fig. 2** Improved thermal conductivity with chip on board LED design.

As the name suggests, an LED chip is basically a diode (pn-junction), designed to allow electrons and holes to recombine to generate photons. This is depicted in Fig. 3 below. The energy levels (and hence wavelengths) of the emitted photons depend on the semiconductor band-gap structures of the chips concerned. The detailed quantum mechanical description of the working principle of LEDs is beyond the scope of this review.

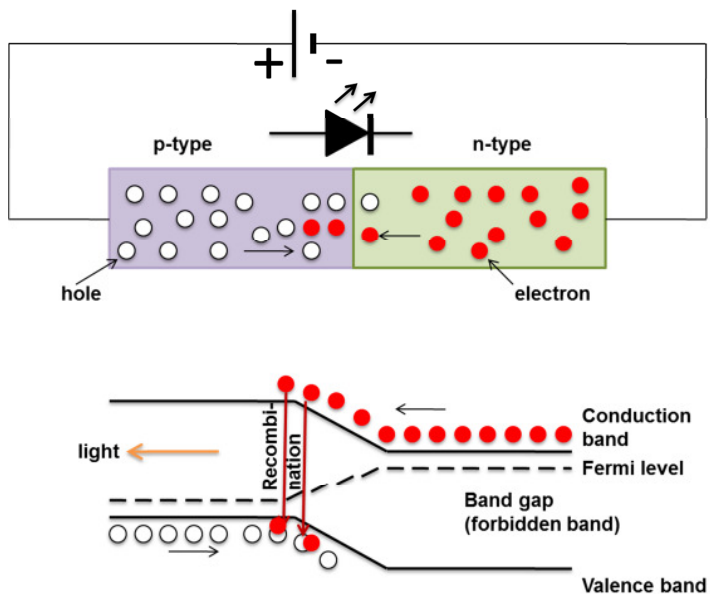

**Fig. 3** Schematics of light emission mechanism inside an LED chip.

As far as efficiency is concerned, note that an incandescent lamp converts <5% of its input electrical energy into light [17] whereas commercial LEDs with >50% efficiency are well known. This clearly indicates the enormous potential of LEDs in energy efficient lighting.

LEDs can provide several benefits to the greenhouse industry [9, 10]:

- Reduction in energy consumption up to 70% compared to traditional light sources.
- Fast switching and steady state operation.
- Simple electronic dimming function.

- Reduction of cable gauge (and hence cost and weight).
- High Relative Quantum Efficiency (RQE): Red light has the highest RQE, meaning it is the most efficient at photosynthesis. Blue light is about 70 to 75% as efficient as red light.
- Stable temperature inside the growth chamber and greenhouse.
- Ability to control spectral composition of blue, green, red, and far-red wavelengths.
- Reduction of heat stress on plants.
- Reduction in watering and ventilation maintenance.
- Lifetime, reliability, and compact size as the major technical advantages over traditional light sources.

**How does light affect plant growth?**

Plants require light throughout their whole life-span from germination to flower and seed production. Three parameters of grow light used in greenhouse industries are relevant: quality, quantity and duration. All three parameters have different effects on plant performance [18]:

*Light quantity (intensity):* Light quantity or intensity is the main parameter which affects photosynthesis, a photochemical reaction within the chloroplasts of plant cells in which light energy is used to convert atmospheric $CO_2$ into carbohydrate.

*Light quality (spectral distribution):* Light quality refers to the spectral distribution of the radiation, i.e. which portion of the emission is in the blue, green, red or other visible or invisible wavelength regions. For photosynthesis, plants respond strongest to red and blue light. Light spectral distribution also has an effect on plant shape, development and flowering (photomorphogenesis).

*Light duration (photoperiod):* Photoperiod mainly affects flowering. Flowering time in plants can be controlled by regulating the photoperiod.

Plants do not absorb all wavelengths of light (solar radiation), they are very selective in absorbing the proper wavelength according to their requirements. The most important part of the light spectrum is 400 to 700 nm which is known as photosynthetically active radiation (PAR), this spectral range corresponds to more or less the visible spectrum of the human eye [19]. Chlorophylls (chlorophyll a and b) play an important role in the photosynthesis but they are not the only chromophores. Plants have other photosynthetic pigments, known as antenna pigments (such as the carotenoids β-carotene, zeaxanthin, lycopene and lutein etc.), which participate in light absorption and play a significant role in photosynthesis (Fig. 4).

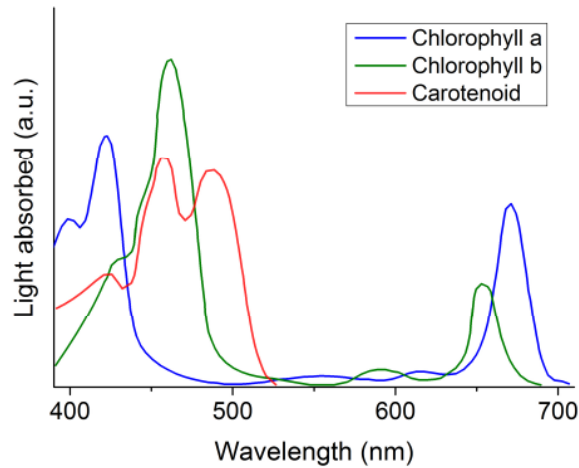

**Fig. 4** Absorption spectrum of chlorophyll and antenna pigments [19].

The solar radiation spectrum mainly consists of three parts: ultraviolet (UV), visible light, and infra-red.

*200-280 nm (ultraviolet C):* This part of the spectrum is harmful to the plant because of its high toxicity. UVC is blocked by the terrestrial ozone layer, so it does not reach the earth's surface.

*280–315 nm (ultraviolet B):* This part is not very harmful but causes plant colors to fade.

*315–380 nm (ultraviolet A):* This range does not have any positive or negative effect on plant growth.

*380–400 nm (ultraviolet A/visible light):* Beginning of visible light spectrum, process of light absorption by plant pigments (chlorophylls and carotenoids) begins.

*400–520 nm (visible light):* Contains violet, blue and green bands. Peak absorption by chlorophylls occurs in this range and it has a strong influence on vegetative growth and photosynthesis.

*520–610 nm (visible light):* This range contains green, yellow and orange bands. This range is less absorbed by the plant pigments and has less influence on vegetative growth and photosynthesis.

*610–720 nm (visible light):* Contains red bands and a large amount of absorption occurs at this range. This band strongly affects the vegetative growth, photosynthesis, flowering and budding.

*720–1000 nm (far-red/infrared):* Germination and flowering is influenced by this range but little absorption occurs at this band.

*>1000 nm (infrared):* All absorption in this region is converted to heat.

Researchers around the world are experimenting with different spectral compositions to optimize the plant growth. A controlled spectrum composition would be much more beneficial for the plants than white light because it would allow to better control the plants' performance such as flowering time, high photosynthetic efficiency, low heat stress etc. LED lighting offers a simple replacement of current light sources (HPS lamps) with better control on spectral composition.

## LEDs as a radiation source for plants

LEDs as a source of plant lighting were used more than 20 years ago when lettuce was grown under red (R) LEDs and blue (B) fluorescent lamps [20]. Several reports have confirmed successful growth of plants under LED illumination [20-23]. Different spectral combinations have been used to study the effect of light on plant growth and development and it has been confirmed that plants show a high degree of physiological and morphological plasticity to changes in spectral quality [24, 25]. Red (610-720 nm) light is required for the development of the photosynthetic apparatus and photosynthesis, whereas blue (400-500 nm) light is also important for the synthesis of chlorophyll, chloroplast development, stomatal opening and photomorphogenesis [26-28]. Several horticultural experiments with potato, radish [29] and lettuce [30] have shown the requirement of blue (400-500 nm) light for higher biomass and leaf area. However, different wavelengths of red (660, 670, 680 and 690 nm) and blue (430, 440, 460 and 475 nm) light might have uneven effects on plants depending on plant species [25, 31, 32]. Far-red LED light (700-725 nm) which is beyond the PAR has been shown to support the plant growth and photosynthesis [30, 31].

As reported by Goins et al. (2001) biomass yield of lettuce increased when the wavelength of red LED emitted light increased from 660 to 690 nm [31]. Stutte et al. (2009) compared the effect of red LED (640 nm) light with far-red LED (730 nm) on the physiology of red leaf lettuce (*Lactua sativa*) [30]. Results showed application of far-red (730 nm) with red (640 nm) caused increase in total biomass and leaf length while anthocyanin and antioxidant potential was suppressed. Mizuno et al. (2011) used red LED (640 nm) light as a sole source and results showed increase in anthocyanin contents in red leaf cabbage (*Brasica olearacea* var. capitata L.) [33]. Addition of far-red (735 nm) to the red (660 nm) LED light on sweet pepper (*Capsicum annum* L.) resulted in taller plants with higher stem biomass than red LEDs alone [34].

Positive effects of blue (400-500 nm) LED light in combination with red LED light on green vegetable growth and nutritional value have been shown in several experiments. Mizuno et al. (2011) and Li et al. (2012) have reported that blue LEDs (440 and 476 nm) used in combination with red LEDs caused higher chlorophyll ratio in Chinese cabbage plants [32, 33]. Goins et al. (1997) reported that wheat (*Triticum aestivum* L., cv. 'USU-Super Dwarf') can complete its life cycle under red LEDs alone but larger plants (higher shoot dry matter) and greater amounts of seed are produced in the presence of red LEDs supplemented with a quantity of blue light [35]. Similar experiments have shown increased nutritional value and enhanced antioxidant status in green vegetables: increased carotenoid [36], vitamin C [32], anthocyanin [30] and polyphenol [37]. Several reports (Table 1) have shown that plant response (growth, flowering time and secondary metabolite) to light quality is species specific. Table 1 contains a summary of various research work carried out on different plant species to study the effect of specific wavelengths (using LEDs as a radiation source) on plants physiology.

**Table 1** Effect of LED lighting on physiology of vegetables

| Plant | Radiation source | Effect on plant physiology | Reference |
|---|---|---|---|
| Indian mustard (*Brassica juncea* L.) Basil (*Ocimum gratissimum* L.) | Red (660 and 635 nm) LEDs with blue (460 nm) | Delay in plant transition to flowering as compared to 460 nm + 635 nm LED combination. | [38] |
| Cabbage (*Brassica olearacea* var. *capitata* L.) | Red (660 nm) LEDs | Increased anthocyanin content. | [33] |
| Baby leaf lettuce (*Lactuca sativa* L. cv. Red Cross) | Red (658 nm) LEDs | Phenolics concentration increased by 6% | [7] |
| Tomato (*Lycopersicum esculentum* L. cv. MomotaroNatsumi) | Red (660 nm) LEDs | Increased tomato yield. | [39] |
| Kale plants (*Brassica olearacea* L. cv Winterbor) | Red (640 nm) LEDs (pretreatment with cool-white light fluorescent lamp) | Lutein and chlorophyll a, b accumulation increased. | [36] |
| Lettuce (*Lactuca sativa*) Green onions (*Allium cepa* L.) | Red (638 nm) LEDs and natural illumination. | Reduction of nitrate content. | [40] |
| White mustard (*Sinapsis alba*), Spinach (*Spinacia oleracea*), Green onions (*Allium cepa*) | Red (638 nm) LEDs with HPS lamp (90 µmol m$^{-2}$ S$^{-1}$), total PPF (photosynthetic photon flux) maintained at 300 µmol m$^{-2}$ S$^{-1}$ | Increased vitamin C content in mustard, spinach and green onions. | [41] |
| Green baby leaf lettuce (*Lactuca sativa* L.) | Red (638 nm) LEDs (210 µmol m$^{-2}$ S$^{-1}$) with HPS lamp (300 µmol m$^{-2}$ S$^{-1}$). | Total phenolics (28.5%), tocopherols (33.5%), sugars (52.5%), and antioxidant capacity (14.5%) increased but vitamin C content | [42] |

| | | decreased. | |
|---|---|---|---|
| Red leaf, green leaf and light green leaf lettuces (*Lactuca sativa* L.) | Red (638 nm) LEDs (300 µmol m$^{-2}$ S$^{-1}$) with HPS lamp (90 µmol m$^{-2}$ S$^{-1}$) | Nitrate concentration in light green leaf lettuce (12.5%) increase but decreased in red (56.2%) and green (20.0%) leaf lettuce | [43] |
| Green leaf 'Lolo Bionda' and red leaf 'Lola Rosa' lettuces (*Lactuca sativa* L.) | Red (638 nm) LEDs (170 µmol m$^{-2}$ S$^{-1}$) with HPS lamp (130 µmol m$^{-2}$ S$^{-1}$) | Total phenolics and α-tocopherol content increased. | [44] |
| Sweet pepper (*Capsicum annuum* L.) | Red (660 nm) and far-red (735 nm) LEDs, total PPF maintained at 300 µmol m$^{-2}$ S$^{-1}$ | Addition of far-red light increased plant height with higher stem biomass. | [34] |
| Red leaf lettuce 'Outeredgeous' (*Lactuca sativa* L.) | Red (640 nm, 300 µmol m$^{-2}$ S$^{-1}$) and far-red (730 nm, 20 µmol m$^{-2}$ S$^{-1}$) LEDs. | Total biomass increased but anthocyanin and antioxidant capacity decreased. | [30] |
| Red leaf lettuce 'Outeredgeous' (*Lactuca sativa* L.) | Red (640 nm, 270 µmol m$^{-2}$ S$^{-1}$) LEDs with blue (440 nm, 30 µmol m$^{-2}$ S$^{-1}$) LEDs. | Anthocyanin content, antioxidant potential and total leaf area increased. | [30] |
| Cherry tomato seedling | Blue LEDs in combination with red and green LEDs, total PPF maintained at 300 µmol m$^{-2}$ S$^{-1}$. | Net photosynthesis and stomatal number per mm$^2$ increased. | [39] |
| Seedlings of cabbage (*Brassica olearacea* var. *capitata* L.) | Blue (470 nm, 50 µmol m$^{-2}$ S$^{-1}$) LEDs alone. | Higher chlorophyll content and promoted petiole elongation. | [33] |
| Chinese cabbage (*Brassica camprestis* L.) | Blue (460 nm, 11% of total radiation) LEDs with red (660 nm) LEDs, total PPF maintained at 80 µmol m$^{-2}$ S$^{-1}$. | Concentration of vitamin C and chlorophyll was increase due to blue LEDs application. | [32] |
| Baby leaf lettuce 'Red Cross' (*Lactuca sativa* L.) | Blue (476 nm, 130 µmol m$^{-2}$ S$^{-1}$) LEDs | Anthocyanin (31%) and carotenoids (12%) increased. | [7] |
| Tomato seedlings 'Reiyo' | Red (660 nm) and blue (450 nm) in different ratios. | Higher Blue/Red ratio (1:0) caused reduction in stem length. | [16] |
| Cucumber 'Bodega' | Blue (455 nm, 7-16 | Application of blue | [45] |
| (*Cucumis sativus*) and tomato 'Trust' (*Lycopersicon esculentum*) | µmol m$^{-2}$ S$^{-1}$) LEDs with HPS lamp (400-520 µmol m$^{-2}$ S$^{-1}$). | LED light with HPS increased total biomass but reduced fruit yield. | |
| Transplant of cucumber 'Mandy F1' | Blue (455 and 470 nm, 15 µmol m$^{-2}$ S$^{-1}$) with HPS lamp (90 µmol m$^{-2}$ S$^{-1}$). | Application of 455 nm resulted in slower growth and development while 470 nm resulted in increased leaf area, fresh and dry biomass. | [46] |
| Red leaf lettuce (*Lactuca sativa* L. cv Banchu Red Fire) | Green 510, 520 and 530 nm LEDs were used, and total PPF was 100, 200 and 300 µmol m$^{-2}$ S$^{-1}$ respectively. | Green LEDs with high PPF (300 µmol m$^{-2}$ S$^{-1}$) was the most effective to enhance lettuce growth. | [37] |
| Tomato 'Magnus F1' Sweet pepper 'Reda' Cucumber | Green (505 and 530 nm, 15 µmol m$^{-2}$ S$^{-1}$) LEDs with HPS lamp (90 µmol m$^{-2}$ S$^{-1}$). | 530 nm showed positive effect on development and photosynthetic pigment accumulation in cucumber only while 505 nm caused increase in leaf area, fresh and dry biomass in tomato and sweet pepper. | [47] |
| Transplant of cucumber 'Mandy F1' | Green (505 and 530 nm, 15 µmol m$^{-2}$ S$^{-1}$) LEDs with HPS lamp (90 µmol m$^{-2}$ S$^{-1}$). | 505 and 530 nm both resulted in increased leaf area, fresh and dry weight. | [46] |

Green light also contributes to the plant growth and development. This has been confirmed by several experiments. Johkan et al. (2012) reported that green LEDs with high PPF (300 µmol m$^{-2}$ S$^{-1}$) are most effective to enhance the growth of lettuce [37]. Novickovas et al. (2012) have found that green (505 and 530 nm) LED light in combination with HPS lamps contributed to the better growth of cucumber [46]. Folta (2004) evaluated the effect of green (525 nm) LED light on germination of *Arabidopsis* seedlings and results showed that seedlings grown under green, red and blue LED light are longer than those grown under red (630 nm) and blue (470 nm) alone [48]. Supplementation of green light enhanced lettuce growth under red and blue LED illumination [49]. Green light alone is not enough to support the growth of plants because it is least absorbed by the plant but when used in combination with red, blue, and far-red, green light will certainly show some important physiological effects. Further investigations are required to study the required level of green photons for optimum plant growth. Experiments with different wavelength of green, red, blue, and far-red lights (provided by LEDs) would be beneficial in determining the species specific optimal wavelength for plant growth. The findings of the light response spectrum studies could be used to design an energy efficient tailored light response spectrum for specific plant species.

## Potential of LEDs in floriculture

Ornamental plants are of high economic importance. Cut flowers and foliage have a wide market around the world. LEDs can also play a key role in floriculture by providing a suitable light spectrum (quality and duration). Light controls the circadian rhythm of plants which means the clocking of plants to day (light) and night (dark) cycles, and this circadian rhythm influences photomorphogenesis. Red and far-red light have been shown to affect photomorphogenesis, thus, the ratio of red and far-red light also plays an important role in regulation of flowering [50, 51]. Flowering in plants is mainly regulated by phytochromes (a group of plant pigments), which occur in two forms: Pr (responds to red light) and Pfr (responds to far-red light). These two pigments (Pr and Pfr) convert back and forth. Pr is converted into Pfr under red light illumination and Pfr into Pr with far-red light (Fig. 5). The active form which triggers flowering is Pfr. Pr is produced naturally in the plant. The ratio of Pr to Pfr is in equilibrium when the plant receives light (day) because Pr is converted into Pfr by red light and Pfr is converted back to Pr by far-red light. Back conversion of Pfr is however also possible in a dark reaction, so it is the night (dark) period which mainly affects the ratio of Pr to Pfr and controls the flowering time in plants [52-55].

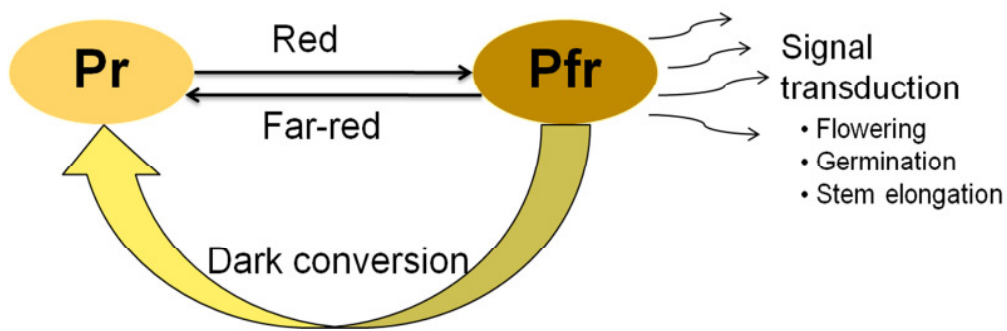

**Fig. 5** Red and far-red light mediated conversion of phytochromes.

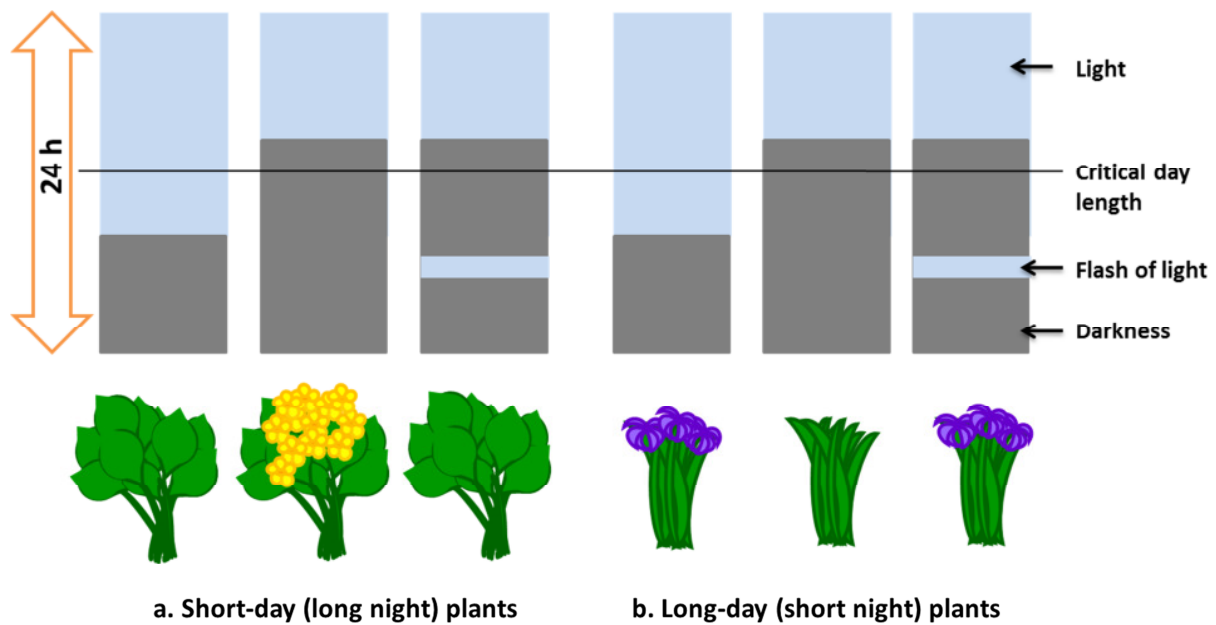

**Fig. 6** Critical night length affects flowering in many plants. Short-day (long-night) plants, such as *Crysanthemum*, flower when the dark period is longer than the critical night length. In contrast, long-day (short-night) plants, such as Iris, flower when the dark period (night) is shorter than critical night length. Flash of light is short duration of light (generally, one to two hours) to interrupt the dark phase [52].

Plants have been divided into two main categories on the basis of day length or photoperiod requirement to flower [44, 52, 53]: Short Day Plants or SDPs (plants flower when the day length is less than their critical night length) and Long Day Plants or LDPs (plants flower when day length is longer than their critical night length, Fig. 6. It is obvious that LDPs require more light (generally more than 14 h of light) to flower and conventional broad-spectrum light sources (incandescent and high pressure sodium lamps) deliver a higher intensity than needed to control flowering and, thus, consume a large amount of energy. LED lighting is an energy efficient option to regulate flowering in long-day ornamental crops because LEDs consume less energy and deliver the specific colors (wavelengths) of light required.

For several long-day plants, addition of far-red light (700-800 nm) to red light (600-700 nm) in order to extend the day length promotes flowering and growth [56]. Meng and Runkle (2014) used 150-Watt incandescent lamps and 14-Watt deep red (DR), white (W) and far-red (FR) LED lamps (developed by Phillips) to study the flowering response in different plants and they found that flowering of bedding plant crops was mostly similar under the Phillips 14-Watt LED (DR+W+FR) lamp as under the conventional 150-Watt incandescent lamps [57]. LEDs (DR+W+FR) are as effective as lamps traditionally used in greenhouses but LEDs are more efficient because they consume only 14 Watt electrical power per lamp. The higher energy efficiency and longer lifetime are the most important advantages of LEDs in floriculture.

**Economic analysis of LEDs in greenhouse industry**

Greenhouse industries have been continuously challenged to provide products (vegetables and flowers) that meet consumers' needs at good market price. In order to control production cost, greenhouse producers must look for the sustainability of resources to meet their operating requirements for the greenhouse cultivation. Heating (to maintain an optimal temperature) and lighting (photoperiod) are the most important cost factors among the various requirements (such as growing media, seeds/cuttings, fertilizers and chemicals etc.). An energy efficient approach can reduce the production cost of green vegetables and ornamental flowers.

The greenhouse market has been increasing very rapidly to supply the required demand of vegetables (especially off-season vegetables) and flowers. On a global scale China is leading with the highest greenhouse cultivation whereas Spain is the major greenhouse vegetable producer in Europe [58]. The results of a horticulture survey published by The Netherlands' ministry of economic affairs, agriculture and innovation [59], showed that tomato, cucumber, field salad and lettuce are the major crops produced by greenhouse industries in Europe. In Germany, all greenhouse industries are growing tomato as their main crop. The economic surveys [59] have reported that 25-35% of production cost for the cultivation of tomatoes is allotted to heating and lighting, and greenhouse industries are looking for new energy efficient approaches to reduce production cost. LEDs can provide the solution for greenhouse lighting with their high energy efficiency and longevity (operating life-time).

Several studies have been carried out to investigate the role of LEDs in commercial greenhouse productions; scientists at Purdue University experimented with LEDs to compare year-round tomato production with supplementing light vs. traditional overhead HPS lighting vs. high intensity red and blue LEDs [60]. The results showed that greenhouse growers can get the same yield of tomato using LEDs which consume 25% energy of the traditional lamps. Similar results have been reported for other crops such as cucumber and lettuce [1]. Traditional lamps (HPS) convert only 30% of the energy into usable light and 30% is lost as heat, whereas LEDs can convert up to 50% and can be optimized for different wavelengths. This shows significant savings in energy, and therefore money, which provides an advantage to the greenhouse industries to compete with production at low cost.

Note that the effects of spatial distribution of light on plant growth can also play a crucial role in the overall productivity [61].

**Operational cost of LEDs and HPS**

As reported by Meng and Runkle recently [57], an HPS lamp of 150 Watt and a 14 Watt LED have a similar effect on the flowering pattern of bedding plants, therefore, use of a 14 Watt LED would be more economical for greenhouse growers.

An estimated calculation of operating cost of LED and HPS for greenhouse growers is presented below:

*Assumptions:*

Average lighting time (during winter) in greenhouse: 16 hours/day

Electricity rate: 0.143 Euro/kWh

*Calculation:*

Electricity consumed by 150-Watt HPS: 2.40 kWh/day

Electricity consumed by 14-Watt LED: 0.22 kWh/day

Annual electricity consumption: 876 kWh (HPS) and 80.3 kWh (LED)

Annual electricity cost in Euro: 125.26 €for HPS and 11.48 € for LED

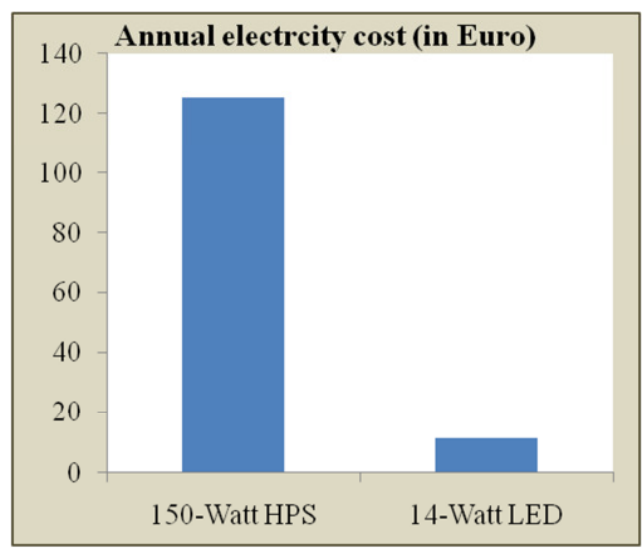

**Fig. 7** Annual electricity cost of a 150-Watt HPS lamp and a 14-Watt LED.

The data (Fig. 7) clearly show that greenhouse growers can reduce the production cost using a proper LED lighting system. High capital cost in LED lighting system is the main factor delaying the penetration of this energy efficient technology into greenhouse industries to date. However, considering the annual electricity cost LEDs will recover the high capital cost and can become a source of profit for greenhouse industries.

**Electricity cost comparison**

LED Grow Master Global has compared the savings of electric energy and reduction in electrical cost between LEDs and high-intensity discharge (HID) lighting, as summarized below. The original web-source can be accessed at (last accessed on 30.05.2014):

http://www.led-grow-master.com/Greenhouse_Cumulative_Cost_LEDs.html

*Assumption:* LEDs are mounted approximately 30 inches above top of plant canopy and high efficiency HPS ballasts. Grow lights are being operated 14 hours per day and 365 days in a year.

*Starting with small grow plot*

| 1' X 3' Area | Power [W] | Energy consumption per year [kWh] | Average cost per kWh [USD] | Cost per year [USD] |
|---|---|---|---|---|
| **1- LGM550** | 9.6 | 49 | 0.10 | 5 |
| **1- 150 w HPS** | 157.5 | 804 | 0.10 | 80 |

*Scale up to large grow plot*

| 12' X 12' Area | Power [W] | Energy consumption per year [kWh] | Average cost per kWh [USD] | Cost per year [USD] |
|---|---|---|---|---|
| **32- LGM550** | 307 | 1,568 | 0.10 | $157 |
| **9- 150 w HPS** | 9450 | 48,289 | 0.10 | $4,829 |

*Cumulative cost comparison over the lifetime of LGM LED grow lights*

LED Grow Master (LGM) Global, the master distributor worldwide for SolarOasis LED grow lights, has compared the cumulative cost factoring initial cost, electricity, disposal and replacement cost.

*LED Assumptions:*

*Lifetime:* LED Grow Master grow lights are rated for 100,000 hours. Utilizing a 14 hour photoperiod, LGM lighting is estimated for a 19 year lifetime.

Initial Cost (*assuming no bulk discount is provided):*80 pieces LGM550 = $23,200 USD

*HPS Assumptions:*

*Lifetime:* HPS bulbs will keep plants productive only as long as the light intensity remains strong. HPS bulbs are generally replaced after 12 months of use if the bulb is used for 12 hours or more a day. Ballasts are calculated for a 6 year lifetime.

*Initial Cost: (assuming no bulk discount is provided):* 9 pieces 1000 Watt HPS bulb, digital ballast, reflector = $5,297 USD

*Bulb Replacement Cost:* Nine pieces 1000 Watt HPS bulb $621 USD

*Disposal Fees:* HPS (classified as hazardous waste) = 9 bulbs = $18 USD

|  | Year 1 | Year 2 | Year 3 | Year 4 | Year 5 | Year 6 | Year 7 | Year 8 |
|---|---|---|---|---|---|---|---|---|
| **80x LGM LEDs** | $28,320 | $23,351 | $23,502 | $23,653 | $23,804 | $23,955 | **$24,106** | $24,257 |
| **9x 1000 watt HPS** | $7,376 | $10,094 | $12,812 | $15,530 | $18,248 | $20,966 | **$28,360** | $31,078 |

|  | Year 9 | Year 10 | Year 11 | Year 12 | Year 13 | Year 14 | Year 15 | Year 16 |
|---|---|---|---|---|---|---|---|---|
| **80x LGM LEDs** | $24,408 | $24,559 | $24,710 | $24,861 | $25,012 | $25,163 | $25,314 | $25,465 |
| **9x 1000 watt HPS** | $33,796 | $36,514 | $39,232 | $41,950 | $49,344 | $52,062 | $54,780 | $57,498 |

Data show that after seven years, cumulative cost of HPS will overpass the LED cost and LEDs will be useful for savings. At the end of 16 years, cumulative cost of HPS will be more than double the amount of LEDs cumulative cost. So, in summary, LEDs require high capital investment but investment will be returned as profit in long operation because LEDs are energy efficient and require less maintenance.

**Conclusions**

This review summarizes the research work done on energy efficient greenhouse lighting with LEDs.

Economic analysis has clearly shown that LEDs can reduce the electricity cost and investment (high capital cost) will be returned as profit in long-term operations in greenhouse industries.
Solid state lighting with LEDs offers high luminous flux and luminance with low radiant heat. LEDs offer the possibility to optimize the light distribution for small and large greenhouses and also in multi-layered farming in greenhouses because LEDs (due to low radiant heat) can be placed close to the plants.
Moreover, optimization of spectral quality to improve plant growth (photosynthetic efficiency, nutritional value and regulation of flowering) and the inherent energy efficiency can reduce power consumption significantly. For example, recent experiments performed on tomato have shown that growers can obtain the same yield with LED lighting in greenhouse with up to 25-30% reduction in production cost compared to conventional lighting.

However, to utilize the full potential of LEDs as a radiation source in greenhouse industries, it is necessary to further investigate the not yet fully understood physiological processes mediating plant responses to LED light. Different light spectra have different effects on plant growth and most studies on the effect of LED radiation on plant physiology have included only red, far-red and blue LED lights as main lighting source. Green light has been considered as photosynthetically inefficient, but even photosynthetically inefficient light can contribute to plant development and growth in orchestration with red and blue light as confirmed by some recent studies. Further investigations are required to understand the roles of green light in regulation of vegetative development, flowering, stem elongation, stomatal opening and plant stature. Research questions such as what specific spectrum, photosynthetic photon flux density and photoperiod are required by different plant species and varieties in different developmental stages have not been conclusively addressed yet, too.

As LED technology provides a lot of flexibility in terms of design of output spectra, adaptation of the lighting conditions to the specific needs of the plants can be achieved. LEDs offer a new energy efficient approach for greenhouse lighting which can reduce the production cost of vegetables and ornamental flowers. However, the potential of this approach is far from being fully explored and more research is required to study effects of LEDs on various vegetables and ornamental plants for large scale industrial applications.

**References**

1. Mitchell CA, Both A, Bourget CM, Kuboto C, Lopez RG, Morrow RC & Runkle S. LEDs: The future of greenhouse lighting. Chronica Horticulture. 2012;55:6-12.

2. Morrow RC. LED lighting in horticulture. Hort Science. 2008;43:1947–1950.

3. Yeh N & Chung JP. High-brightness LEDs – energy efficient lighting sources and their potential in indoor plant cultivation. Renew Sust Energ Rev. 2009;13:2175–2180.

4. Tennessen DJ, Singsaas EL & Sharkey TD. Light-emitting diodes as a light source for photosynthesis research. Photosynth Res. 1994;39:85–92.

5. Barta DJ, Tibbits TW, Bula RJ & Morrow, RC. Evaluation of light emitting diode characteristics for a space-based plant irradiation source. Adv Space Res. 1992;12:141–9.

6. Olle M & Virsile A. The effect of light-emitting diode lighting on greenhouse plant growth and quality. Agric Food Sci. 2013;22:223-234.

7. Li Q & Kubota C. Effects of supplemental light quality on growth and phytochemicals of baby leaf lettuce. Environ Exp Bot. 2009;67:59–64.

8. Lin KH, Huang MY, Huang WD, Hsu MH, Yang ZW & Yang CM. The effects of red, blue, and white light-emitting diodes on the growth, development, and edible quality of hidroponically grown lettuce (*Lactuca sativa* L. var. capitata). SciHortic-Amsterdam. 2013;150:86–91.

9. Massa GD, Kim HH, Wheeler RM & Mitchell CA. Plant productivity in response to LED lighting. Hort Science. 2008;43:1951–1956.

10. Vänninen I, Pinto DM, Nissinen AI, Johansen NS & Shipp L. In the light of new greenhouse technologies: Plant-mediates effects of artificial lighting on arthropods and tritrophic interactions. Ann Appl Biol. 2010;157:393–414.

11. Bourget CM. An introduction to light-emitting diodes. Hort Science. 2008;43:1944–1946.

12. Brumfield R. Dealing with rising energy costs. GPN. 2007;17:24-31.

13. Langton A, Plackett C & Kitchener H. Energy saving in poinsettia production. Horticultural Development Council Fact sheet. 2006;7:1-12.

14. Opdam JG, Schoonderbeek GG, Heller EB & Gelder A. Closed greenhouse: a starting point for sustainable entrepreneurship in horticulture. Acta Hort. 2005;691:517-524.

15. Ieperen VW & Trouwborst G. The Application of LEDs as Assimilation Light Source in Greenhouse Horticulture: a Simulation Study. Acta Hort. 2008;33:1407-1414.

16. Nanya K, Ishigami Y, Hikosaka S & Goto E. Effects of blue and red light on stem elongation and flowering of tomato seedlings. Acta Hort. 2012;956:261–266.

17. Keefe TJ. "The Nature of Light".Archived from the original on 2012-07-24. Retrieved 2007-11-05 Tower Hall Funabori, Tokyo, Japan.

18. Nishio JL. Why are higher plants green? Evolution of the higher plant photosynthetic pigment complement. Plant Cell Environ. 2000;23:539–548.

19. Chen P. Chlorophyll and other photosentives. In: LED grow lights, absorption spectrum for plant photosensitive pigments. http://www.ledgrowlightshq.co.uk/chlorophyll-plant-pigments/. Accessed 12 March 2014.

20. Bula RJ, Morrow RC, Tibbits TW, Barta RW, Ignatius RW & Martin TS. Light emitting diodes as a radiation source for plants. Hort Science.1991;26:203–205.

21. Tanaka Y, Kimata K & Aiba H. A novel regulatory role of glucose transporter of *Escherichia coli*: membrane sequestration of a global repressor Mic. EMBO J. 2000;19:5344-5352.

22. Tripathy BC & Brown CS. Root-shoot interaction in the greening of wheat seedlings grown under red light. Plant Physiol. 1995;107:407–411.

23. Yanagi T & Okamoto K. Utilization of super-bright light emitting diodes as an artificial light source for plant growth. Acta Hort. 1997;418:223-228.

24. Barreiro R, Guiamet JJ, Beltrano J & Montaldi ER. Regulation of the photosynthetic capacity of primary bean leaves by the red: far-red ratio and photosynthetic photon flux density of incident light. Physiol. Plant. 1992;85:97–101.

25. Sims DA & Pearcy RW. Response of leaf anatomy and photosynthetic capacity in Alocasiamacrorrhiza (Araceae) to a transfer from low to high light. Am J Bot. 1992;79:449–455.

26. Akoyunoglou G & Anni H. Blue light effect on chloroplast development in higher plants. In: Senger H. (ed.), Blue Light Effects in Biological Systems. Springer-Verlag, Berlin: 1984. pp. 397–406.

27. Saebo A, Krekling T & Appelgren M. Light quality affects photosynthesis and leaf anatomy of brich plantlets in vitro.Plant Cell Tiss Org. 1995;41:177–185.

28. Senger H. The effect of blue light on plants and microorganisms. Phytochem Photobiol. 1982;35:911–920.

29. Yorio NC, Goins GD, Kagie HR, Wheeler RM & Sager JC. Improving spinach, radish and lettuce growth under red light emitting didoes (LEDs) with blue light supplementation. Hort Science. 2001;36:380–383.

30. Stutte GW, Edney S & Skerritt T. Photoregulation of bioprotectant content of red leaf lettuce with light-emitting diodes. Hort Science. 2009;44:79–82.

31. Goins GD, Ruffe LM, Cranston NA, Yorio NC, Wheeler RM & Sager JC. Salad crop production under different wavelengths of red light-emitting diodes (LEDs). SAE Technical Paper, 31st International Conference on Environmental Systems, July 9–12, 2001, Orlando, Florida, USA: 2001. p. 1–9.

32. Li H, Tang C, Xu Z, Liu X & Han X. Effects of different light sources on the growth of non-heading chinese cabbage (*Brassica campestris* L.). J Agr Sci. 2012;4:262–273.

33. Mizuno T, Amaki W & Watanabe H. Effects of monochromatic light irradiation by LED on the growth and anthocyanin contents in laves of cabbage seedlings. Acta Horticulturae. 2011;907:179–184.

34. Brown C, Shuerger AC & Sager JC. Growth and photomorphogenesis of pepper plants under red light-emitting diodes with supplemental blue or far-red lighting. J Am SocHortic Sci. 1995;120:808–813.

35. Goins GD, Yorio NC, Sanwo MM & Brown CS. Photomorphogenesis, photosynthesis and seed yield of wheat plants grown under red light-emitting diodes (LEDs) with and without supplemental blue lighting. J Exp Bot. 1997;48:1407–1413.

36. Lefsrud MG, Kopsell DA & Sams CE. Irradiance from distinct wavelength light-emitting diodes affect secondary metabolites in kale. Hort Science. 2008;43:2243–2244.

37. Johkan M, Shoji K, Goto F, Hahida S & Yoshihara T. Effect of green light wavelength and intensity on photomorphogenesis and photosynthesis in *Lactuca sativa*. Environ Exp Bot. 2012;75:128–133.

38. Tarakanov I, Yakovleva O, Konovalova I, Paliutina G & Anisimov A. Light-emitting diodes: on the way to combinatorial lighting technologies for basic research and crop production. ActaHorticulturae. 2012;956:171–178.

39. Lu N, Maruo T, Johkan M, Hohjo M, Tsukakoshi S, Ito Y, Ichimura T & Shinohara Y. Effects of supplemental lighting with light-emitting diodes (LEDs) on tomato yield and quality of single-truss tomato plants grown at high planting density. Environ Control Biol. 2012;50:63–74.

40. Samuolienė G, Urbonavičiūtė A, Duchovskis P, Bliznikas Z, Vitta P & Žukauskas A. Decrease in nitrate concentration in leafy vegetables under a solid-state illuminator. Hort Science. 2009;44:1857–1860.

41. Bliznikas Z, Žukauskas A, Samuolienė G, Viršilė A, Brazaitytė A, Jankauskienė J, Duchovskis P & Novičkovas A. Effect of supplementary pre-harvest LED lighting on the antioxidant and nutritional properties of green vegetables. Acta Hort. 2012;939:85–91.

42. Samuolienė G, Sirtautas R, Brazaitytė A, Viršilė A & Duchovskis P. Supplementary red-LED lighting and the changes in phytochemical content of two baby leaf lettuce varieties during three seasons. J Food Agric Environ. 2012a;10:701 – 706.

43. Samuolienė G, Brazaitytė A, Sirtautas R, Novičkovas A & Duchovskis P. Supplementary red-LED lighting affects phytochemicals and nitrate of baby leaf lettuce. J Food Agric Environ. 2011;9:271–274.

44. Žukauskas A, Bliznikas Z, Breivė K, Novičkovas A, Samuolienė G, Urbonavičiūtė A, Brazaitytė A, Jankauskienė J & Duchovskis P. Effect of supplementary pre-harvest LED lighting on the antioxidant properties of lettuce cultivars. Acta Hort. 2011;907:87–90.

45. Ménard C, Dorais M, Hovi T & Gosselin A. Developmental and physiological responses of tomato and cucumber to additional blue light. Acta Hort. 2006;711:291–296.

46. Novičkovas A, Brazaitytė A, Duchovskis P, Jankauskienė J, Samuolienė G, Viršilė A, Sirtautas R, Bliznikas Z & Žukauskas A. Solid-state lamps (LEDs) for the short-wavelength supplementary lighting in greenhouses: experimental results with cucumber. Acta Hort. 2012;927:723–730.

47. Samuolienė G, Brazaitytė A, Duchovskis P, Viršilė A, Jankauskienė J, Sirtautas R, Novičkovas A, Sakalauskienė S & Sakalauskaitė,J. Cultivation of vegetable transplants using solid-state lamps for the short-wavelength supplementary lighting in greenhouses. Acta Hort. 2012c ;952:885–892.

48. Folta KM. Green light stimulates early stem elongation, antagonizing light-mediated growth inhibition. Plant Physiol. 2004;135:1407–1416.

49. Kim HH, Goins GD, Wheeler RM & Sager JC. Green- light supplementation for enhanced lettuce growth under red and blue light-emitting diodes. Hort Science. 2004;39:1617–1622.

50. Simpson GG & Dean C. *Arabidopsis*, the Rosetta stone of flowering time? Science. 2002; 296:285–289.

51.Yanovsky MJ & Kay SA. Molecular basis of seasonal time measurement in *Arabidopsis*. Nature. 2002;419:308–312.

52. Downs RJ & Thomas JF. Phytochrome regulation of flowering in the long-day plant, Hyoscyamusniger. Plant Physiol. 1982;70:898–900.

53. Evans LT. Inflorescence initiation in *Loliumtemu lentum* L. XIV. The role of phytochrome in long day induction. Austral. J. Plant Physiol. 1976;3:207–217.

54. Shinomura T, Uchida K & Furuya M. Elementary processes of photoperception by phytochrome A for high-irradiance response of hypocotyl elongation in *Arabidopsis*. Plant Physiol. 2000;122:147–156.

55. Smith H. Light quality, photoperception, and plant strategy. Annu Rev Plant Physiol. 1982;33:481–518.

56. Runkle ES & Heins DR. Specific functions of red, far-red and blue lights in flowering and stem extension of long-day plants. J Amer Soc. Hort Sci. 2001;126:275–282.

57. Meng Q & Runkle ES. Control flowering with LEDs. Lighting Research.Growers Talk 62. http://www.ballpublishing.com/GrowerTalks/ViewArticle.aspx?articleid=20604 Accessed 15 Feb 2014.

58. Gomez C, Morrow RC, Bourget CM, Massa GD & Mitchell CA. Comparison of intracanopy light-emitting diode towers and overhead high-pressure sodium lamps for supplemental lighting of greenhouse-grown tomatoes. Hort Technology. 2013;23:93–98.

59. Voss J. Market special: greenhouse farming in Germany. The ministry of Economics Affairs, Agriculture and Innovation, NL, EVD International. 2011. http://duitsland.nlambassade.org/binaries/content/assets/postenweb/d/duitsland/ambassade-berlijn/zaken-doen/20110507-marktverkenning-greenhouse-farming-germany.pdf Accessed 16 Feb 2014

60. Kacira, M. Greenhouse Production in US: Status, Challenges, and Opportunities. Presented at CIGR 2011 conference on Sustainable Bioproduction WEF 2011, September 19-23, 2011.

61. Nelson AJ & Bugbee B. 2013. Supplemental greenhouse lighting: Return on investments for LED and HPS fixtures. http://cpl.usu.edu/files/publications/factsheet/pub__4338884.pdf